\documentclass[aps,showpacs,twocolumn,amsmath,amssymb,superscriptaddress]{revtex4-1}
\usepackage{graphicx}
\usepackage{epstopdf}
\usepackage{bm}
\usepackage{subfigure}
\usepackage{sidecap}
\usepackage{times}
\usepackage{graphicx}
\usepackage{epstopdf}
\usepackage{amssymb}
\usepackage{amsmath}
\usepackage{dsfont,amsthm,amsbsy}
\usepackage{amssymb}
\usepackage{amsmath}
\usepackage{bbm}
\usepackage{graphicx}
\usepackage{epstopdf}
\usepackage{subfigure}
\usepackage{natbib}
\usepackage{epsfig}
\usepackage{amsfonts}
\usepackage{mathrsfs}
\usepackage{sidecap}
\usepackage{lipsum}
\usepackage{xcolor}
\usepackage[toc,page,title,titletoc,header]{appendix}
\usepackage[colorlinks,linkcolor=blue,citecolor=blue,anchorcolor=blue, urlcolor=blue]{hyperref}
\usepackage{hyperref}
\usepackage{resizegather}
\usepackage{float}
\usepackage{mathbbol}
\usepackage[normalem]{ulem}
\usepackage{cancel}
\usepackage{upgreek}
\usepackage{graphics,dcolumn,bm}
\usepackage{rotate,color}
\usepackage{times}
\newcommand\redsout{\bgroup\markoverwith{\textcolor{red}{\rule[0.5ex]{2pt}{0.4pt}}}\ULon}

\newcommand{\bl}{\begin{aligned}}
\newcommand{\el}{\end{aligned}}
\def\be{\begin{equation}}
\def\ee{\end{equation}}

\def\bi{\begin{itemize}}
\def\ei{\end{itemize}}
\def\bn{\begin{enumerate}}
\def\en{\end{enumerate}}
\def\bea{\begin{eqnarray}}
\def\eea{\end{eqnarray}}
\def\no{\nonumber}
\def\ba{\begin{array}}
\def\ea{\end{array}}
\def\bd{\begin{displaymath}}
\def\ed{\end{displaymath}}

\graphicspath{{Figures/}}
\begin{document}

\title{Floquet dynamical phase transition and entanglement spectrum} 

\author{R. Jafari}
\email{rohollah.jafari@gmail.com}
\affiliation{Department of Physics, Institute for Advanced Studies in Basic Sciences (IASBS), Zanjan 45137-66731, Iran}
\affiliation{Department of Physics, University of Gothenburg, SE 412 96 Gothenburg, Sweden}
\affiliation{Beijing Computational Science Research Center, Beijing 100094, China}
\author{Alireza Akbari}
\email{akbari@postech.ac.kr}
\affiliation{Max Planck Institute for the  Chemical Physics of Solids, D-01187 Dresden, Germany}
\affiliation{Max Planck POSTECH Center for Complex Phase Materials, and Department of Physics, POSTECH, Pohang, Gyeongbuk 790-784, Korea}
\affiliation{Department of Physics, Institute for Advanced Studies in Basic Sciences (IASBS), Zanjan 45137-66731, Iran}

\begin{abstract}
We explore both pure and mixed states Floquet dynamical quantum phase transitions (FDQFTs) in the one-dimensional
\textit{p}-wave superconductor with a time-driven pairing phase. In the Fourier space, the model is recast to the non-interacting quasi-spins subjected to a time-dependent effective magnetic field.
We show that FDQFTs occur within a range of driving frequency without resorting to any quenches.
Moreover, FDQFTs appear in the region where quasi-spins are in the resonance regime.
In the resonance regime, the population completely cycles the population between the spin down and up states. Additionally, we study the conditions for the appearance of FDQFTs using the entanglement spectrum and purity entanglement measure. Our results imply that the entanglement spectrum can truly capture the resonance regime where FDQFTs occur.
Particularly, the dynamical topological region results in the degeneracy of the entanglement spectrum.
It is shown that the boundary of the driven frequency range, over which the system reveals FDQFTs, signaled by the purity entanglement measure.
\end{abstract}

\maketitle

\section{Introduction}
Quantum phase transition (QPT), in a similar fashion as a classical phase transition, is one of the most intriguing research topics in condensed-matter physics~\cite{sachdevbook}.
It is characterized by signaling nonanalytic behaviors in some physical properties of the system~\cite{Vojta2003} and is often accompanied by a divergence in some correlation functions. But, the quantum systems possess additional ``quantum correlations"
which do not exist in classical counterparts~\cite{nielsen2000}.
Consequently, quantum correlations could be useful to investigate the quantum phase transition~\cite{Campbell2013,Preskill,Jafari2020a,Mahdavifar2017,Mukherjee2016,Jafari2015,Marzolino2014,Mishra2018}.

Entanglement is a type of quantum correlation first signified by Schr\"{o}dinger in 1935~\cite{schrodinger1935} as  a particular
feature of quantum mechanics. As a direct measure of quantum correlations, it displays nonanalytic behavior such as discontinuity at the quantum critical points~\cite{osterloh2002,Wu2004,Jafari2008}. In the past decade, the subject of several pieces of research was to explore the behavior of entanglement near and at the quantum critical point for different spin models~\cite{Vidal2003,osterloh2002,Osborne2002,Verstraete2004} as well as itinerant systems~\cite{zanardi2002,Gu2004,Anfossi2007}. Furthermore, purity entanglement measure~\cite{Barnum2003,Barnum2004,Somma2004,Batle2015} and entanglement spectrum (ES)~\cite{Gong2018,Li2008,Chang2020,Stojanovi2020,Chang2020,Lu2019} introduced for quantifying the characteristics of quantum entanglement in many-body systems.

Recently, a new research area of quantum phase transition introduced in nonequilibrium quantum systems, named dynamical quantum phase transitions (DQPTs) as a counterpart of equilibrium thermal phase transitions~\cite{Heyl2013,heyl2018dynamical}.
The notion of DQPT emanates from the similarity between the equilibrium partition function of a system and Loschmidt amplitude, which measures the overlap between an initial state and its time-evolved one~\cite{Heyl2013,heyl2018dynamical,jafari2019quench,Jafari2017,Divakaran2013,Guo2020,Najafi2018,Najafi2019,Yan2020,Zache2019,
Mukherjee2019,Wang2017,Zhang2016,Zhang2016b,Serbyn2017,Jafari2016}.
While the equilibrium phase transition is characterized by non-analyticities in the thermal free energy, the DQPT is signaled by the nonanalytical behavior of dynamical free energy, in which the real-time plays the role of the control parameter~\cite{andraschko2014dynamical,sedlmayr2018fate,vajna2015topological,Karrasch2013,vajna2014disentangling,jafari2019dynamical}.
Further, analogous to order parameters at equilibrium quantum phase transition, a dynamical topological order parameter 
is proposed to capture DQPTs~\cite{budich2016dynamical}.
It is quantized and its unit magnitude jumps at the time of DQPT reveals the topological characteristic feature of DQPT~\cite{budich2016dynamical,Bhattacharjee2018,Dutta2017,sharma2014loschmidt}.

More recently, significant theoretical~\cite{Divakaran2016,Sharma2016b,bhattacharya2017emergent,weidinger2017dynamical,zhou2018dynamical,canovi2014first,
hickey2014dynamical,vzunkovivc2018dynamical,Zhou2019,Mera2018,Khatun2019,Sedlmayr2018,Sedlmayr2020,sharma2015quenches,
Srivastav2019,Abdi2019,Cao2020,Bhattacharyya2020,Jafari2020,Rylands2020,Hu2020,Pastori2020,Kyaw2020,Mishra2020,Puebla2020,Ding2020} and experimental~\cite{flaschner2018observation,jurcevic2017direct,martinez2016real,guo2019observation,wang2019simulating,Nie2020,Tian2020}
endeavors have focussed on DQPTs.
On the theoretical front, most of researches are devoted to study the DQPTs of both slow and sudden quantum quenches of the Hamiltonian.
Furthermore, few works attempt to provide a link between sudden quench DQPTs and entanglement~\cite{Sun2018,Nicola2020},
entanglement entropy~\cite{jurcevic2017direct,schmitt2015dynamical,Canovi2014}, and entanglement spectrum~\cite{Canovi2014,Torlai2014,Surace2020,Su2020}.
Lately, time-periodic driving and the corresponding Floquet theory has been attracted great attention~\cite{kosior2018dynamical,kosior2018dynamicalb,yang2019floquet,Zamani2020}.
The study of time-periodically driven closed quantum systems in the context of the Floquet theory is one of the most attractive areas of developing non-equilibrium research.
Despite considerable investigation many aspects of DQPTs~\cite{Divakaran2016,Sharma2016b,bhattacharya2017emergent,weidinger2017dynamical,zhou2018dynamical,canovi2014first,
hickey2014dynamical,vzunkovivc2018dynamical,Zhou2019,Mera2018,Khatun2019,Sedlmayr2018,Sedlmayr2020,sharma2015quenches,
Srivastav2019,Abdi2019,Cao2020,Bhattacharyya2020,Jafari2020,Ding2020,Rylands2020,Hu2020,Pastori2020},
comparatively, little attention has been directed toward Floquet DQPTs~\cite{kosior2018dynamical,kosior2018dynamicalb,yang2019floquet,Zamani2020}.
To make progress, more studies are needed, specifically, the exactly solvable models play an important role.

The main aim of this study is to find the connection between Floquet DQPTs and purity entanglement measure and entanglement spectrum. Such contributions can bring several new realizations to the subject.
Here, we study analytically both pure and mixed states  Floquet dynamical quantum phase transitions (FDQFTs) in the one-dimensional \textit{p}-wave superconductor with a time-driven pairing phase. We show that FDQPTs occur without requiring any quenches at the region, where the population between spin down and up states is completely cycled.
We also investigate the conditions for the appearance of DQPTs using entanglement spectrum and purity entanglement measures.
The range of driving frequency over that the system is dynamically topological and the dynamical topological QPT (DTQPT) arises, as well as, the region where DTQPT happens, signaled by the degeneracy of the entanglement spectrum, can truly be detected by the entanglement spectrum and the purity entanglement measures.


\section{Theoretical Model}

The Hamiltonian of one dimensional \textit{p}-wave superconductor  with time dependent pairing phase (magnetic flux) is given as~\cite{Kitaev2001}
%
\begin{equation}
\bl
{\cal H}=\sum_{j=1}^{N}
\Big[
\Big(\frac{{\cal W}}{2}c_{j}^{\dagger}c_{j+1}-\frac{\Delta}{2}
e^{-{\it i}\theta(t)}c_{j}^{\dagger}c_{j+1}^{\dagger}+H.C\Big)
-\mu(c^{\dagger}_{j} c_{j}-\frac{1}{2})
\label{eq1}
\Big]
\el
\end{equation}
%
where $c_{j}$ ($c_{j}^{\dagger}$) is the fermion creation (annihilation) operator, $N$ is the number of lattice sites and $\mu$ is the chemical
potential. The hopping and pairing amplitudes are $w$ and $\Delta$, respectively.
The phase factor $\theta(t)$ in the pairing terms is the vector potential, interpreting as an Aharonov-Bohm flux $\Phi(t)=N\theta(t)$
piercing the ring~\cite{Nakagawa2016}.
This model can be mapped to the periodically time-dependent
extended XY spin model via a Jordan-Wigner transformation.
To diagonalize the fermionic Hamiltonian in Eq.~(\ref{eq1}) we perform a Fourier transform,
%
$
c_{j} = \frac{1}{\sqrt{N}} \sum_{k} c_{k} e^{{\it i} kj},
$
and
$ c_{j}^{\dagger}
= \frac{1}{\sqrt{N}} \sum_{k} c_{j}^{\dagger} e^{-{\it i} kj}.
$
%
Considering  antiperiodic boundary conditions ($c_{j+N}=-c_{j}$),
 results the wave number $k=(2p-1)\pi/N$, where $p$ runs from $1$ to $N/2$.
Introducing fermionic two-component $\Gamma^{\dagger}_{k}=(c_{k}^{\dagger},~c_{-k})$, the Hamiltonian of Eq.~(\ref{eq1})
can be written as the sum of $N$ non-interacting terms
%
\begin{equation}
\label{eq2}
{\cal H}=\sum_{k}\Gamma^{\dagger}_{k}H_{k}(t)\Gamma_{k},
\end{equation}
%
where the Bloch Hamiltonian $H_{k}(t)$ is defined  as
%
\begin{equation}
\bl
\no
\label{eq3}
H_{k}(t) =
\frac{1}{2}
\Big[h_{xy}(k)
[
\sin(\omega t)\sigma_{x}
\!-\!
\cos(\omega t)\sigma_{y}
]
+
h_{z}(k)\sigma_{z}\Big],
\el
\\
\end{equation}
%
with $h_{z}(k)={\cal W}\cos(k)-\mu$,~and $h_{xy}(k)=\Delta\sin(k)$, where $\sigma_{\alpha=0,x,y,z}$ are Pauli matrices.
%
We should mention that, Eq.~(\ref{eq2}) expresses that the Hamiltonian of interacting fermions system, Eq.~(\ref{eq1}),
mapped to the sum of noninteracting quasi-spins imposed by the time-dependent effective magnetic field.
The single particle quasi-spin Hamiltonian $H_{k}(t)$, is exactly the Schwinger-Rabi model of a spin in a rotating magnetic field~\cite{Schwinger1937}.
The exact solution to the time-dependent Schr\"{o}dinger equation, ${\it i}\frac{d}{dt}|\psi(k,t)\rangle=H_{k}(t)|\psi(k,t)\rangle$,
is achieved by going to the rotating frame given by the periodic unitary transformation~\cite{Rodriguez2018}, 
%
\begin{equation}
\label{eq4}
U_{R}(t)=
e^{{\it i}\omega( \sigma_{0} -\sigma_{z})t/2}
=
\left(
           \begin{array}{cc}
             1 & 0 \\
             0 & e^{{\it i}\omega t} \\
           \end{array}
         \right),
\end{equation}
%
to obtain  the time-independent Flouquet Hamiltonian, $H_{F}$, as
%
\begin{equation}
\bl
\label{eq5}
H_{k}^{F}=
&
\;
U_{R}^{\dagger}(t)H_{k}(t)U_{R}(t)-{\it i}U_{R}^{\dagger}(t)\frac{dU_{R}(t)}{dt}
\\
=
&-\frac{1}{2}\Big[h_{xy}(k)\sigma_{y}-
[
h_{z}(k)-\omega
]
\sigma_{z}-\omega \sigma_{0} 
\Big].
\el
\end{equation}
%
The eigenvalues and eigenvectors of the Floquet Hamiltonian $H_{k}^{F}$ are given by
%
\begin{equation}
\bl
\label{eq6}
\varepsilon^{\pm}_{k}=
&\frac{1}{2}
\Big[
\omega
\pm
\sqrt{h_{xy}^{2}(k)+
[
h_{z}(k)-\omega
]^{2}
}
\Big],
\el
\end{equation}
and
\begin{equation}
\bl
|\chi^{\pm}_{k}\rangle=
&\frac{1}{N_{k}}
\Big[
h_{xy}(k)
| \mp \rangle
+{\it i}
\eta^{}_z(k)
| \pm \rangle
\Big],
%
%
%
%
\el
\end{equation}
%
respectively. Here we define
$$N_{k}=\sqrt{
\eta^{2}_z(k)
+h_{xy}^{2}(k)};
\;\;\;
\;\;\;
\;\;\;
\\
\eta^{}_z(k)=
[h_{z}(k)-2\varepsilon^{-}_{k}
],
$$
and
$| \pm \rangle$ 
are the eigenstates of $\sigma_{z}$.
In the original frame, the Floquet states of the Hamiltonian $H_{k}(t)$
is given by
%
\begin{equation}
\label{eq7}
|\psi^{\pm}_{k}(t)\rangle=U_{R}(t)e^{-{\it i}H^{F}_{k}t}|\chi^{\pm}_{k}\rangle=e^{-{\it i}\varepsilon^{\pm}_{k}t}U_{R}(t)|\chi^{\pm}_{k}\rangle.
\end{equation}
%
Consequently, the initial and time evolved ground states of the original Hamiltonian are obtained  as follows
%
\begin{equation}
\bl
\label{eq8}
|\psi^{-}(t)\rangle
=\,
&\Pi_{k}|\psi^{-}_{k}(t)\rangle=
\Pi_{k} e^{-{\it i}
\varepsilon^{-}_{k}t}U_{R}(t)|\chi^{-}_{k}\rangle,\\
|\psi^{-}(0)\rangle
=\,
&\Pi_{k}|\chi^{-}_{k}\rangle.
\el
\end{equation}
%
Moreover, in fermion language the ground state of the proposed time-dependent Hamiltonian Eq.~(\ref{eq1})
is given as
%
\bea 
\label{eq9}
|\psi^{-}(t)\rangle=\prod_{k>0}^{}
\left[
 u_k(t,\omega) + v_k(t,\omega)c_k^{\dagger}c_{-k}^{\dagger}
 \right]
 \lvert0\rangle,
\eea 
%
where
$u_k(t,\omega)=h_{xy} (k) e^{-{\it i}\varepsilon^{-}_{k}t}/N_{k}$ and
$
v_k(t,\omega)={\it i}
\eta^{}_z(k)
e^{-{\it i}\varepsilon^{-}_{k}t}e^{{\it i}\omega t}
 /N_{k}
 $.
If we assume the system initially prepared in $|-\rangle$ state at $t=0$,
the probability of the transition from the state $|-\rangle$  to the state $|+\rangle$  (spin flip probability)
is given as
%
\begin{equation}
\label{eq10}
P_{f}=
\frac{h_{xy}^{2}}{
\Omega_{R}
}\sin^{2}(\frac{\Omega_{R}t}{2})
;
\;\;\;
\;\;\;
\Omega_{R}=\sqrt{h_{xy}^{2}(k)+
[h_{z}(k)-\omega
]^{2}}.
\end{equation}
%
Note that whenever $h_{z}=\omega$,  the spin flip (Rabi transition) probability
can become maximum possible value $1$.
In such a resonance situation, the period of oscillation $T_{R}=2\pi/\Omega_{R}$, is
different from the driving period~\cite{Schwinger1937}. In the other words, the population at resonance completely
cycles the population between the two spin down and up states, while for
$h_{z}\neq\omega$,
the down state $|-\rangle$ is never completely depopulated.

%
\begin{figure*}[t]
\centerline{\includegraphics[width=0.71\textwidth]{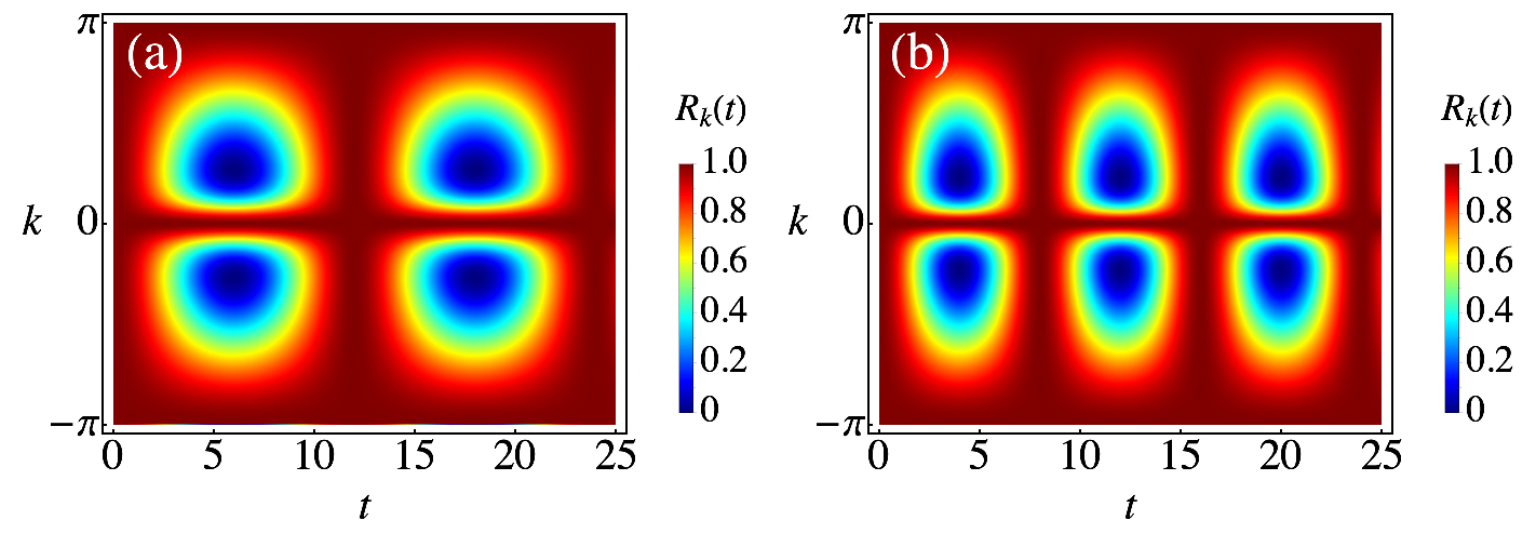}
\hspace{0.2cm}
\includegraphics[width=0.53\columnwidth]{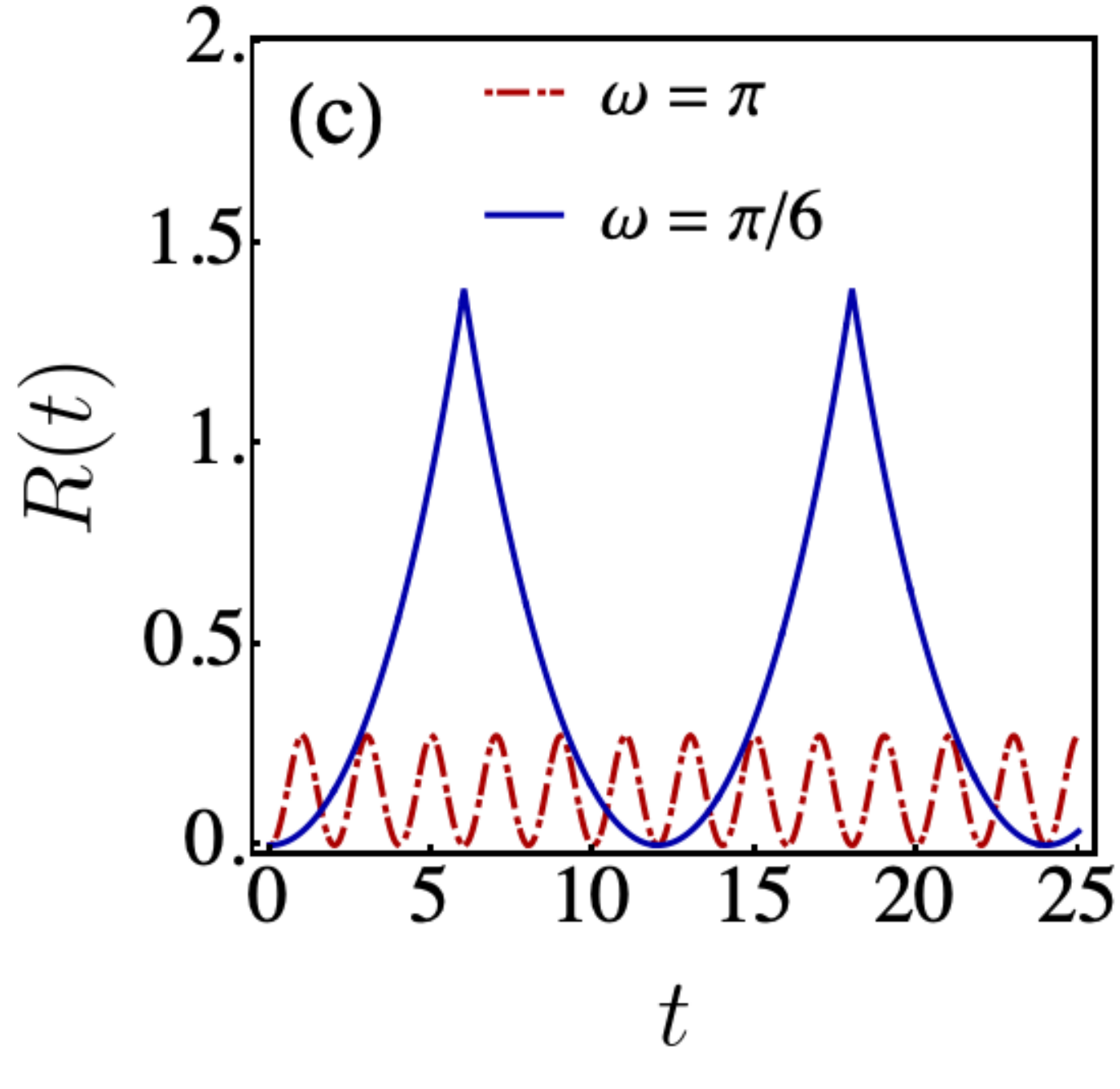}}
\vspace{-0.35cm}
\caption{(Color online) The density plot of Loschmidt echo versus $t$ and $k$ for (a) $\omega=\pi/6$, and (b) $\pi/4$.
(c) The dynamical free energy $g(t)$ versus $t$ for $\omega=\pi/6$ (solid line) and $\omega=\pi$ (dotted line). 
Here we set ${\cal W}=\Delta=\pi$ and $\mu=\pi/2$.}
\label{fig1}
\end{figure*}
%

\section{Dynamical quantum phase transition}
As mentioned, the concept of a DQPT extracted from the analogy between
the partition function of an equilibrium system
$Z(\beta)={\rm Tr} [ e^{-\beta {\cal H}}]$
and the boundary quantum partition function $Z(z)=\langle\psi_{0}|e^{-z {\cal H}}|\psi_{0}\rangle$
with $|\psi_{0}\rangle$ a boundary state and $z \in \mathds{C}$.
When $z=it$, the boundary quantum partition function becomes equivalent to a Loschmidt amplitude (LA),
${\cal L}(t)=\langle\psi_{0}|e^{-{\it i}  {\cal H}t}|\psi_{0}\rangle$, denoting the overlap between the initial state $|\psi_{0}\rangle$
and the time-evolved one $|\psi_{0}(t)\rangle$~\cite{Heyl2013}.
Heyl {\em et al.}~\cite{Heyl2013} showed  that, similar to the thermal free
energy, a dynamical free energy can be defined as
$$g(t)=-\frac{1}{2\pi}\int_{-\pi}^{\pi} dk \ln|{\cal L}_{k}(t)|^{2},
$$
where the real time $t$, plays the role of the control parameter.
DQPTs are simply signaled by non-analytical behavior of $g(t)$ as a function of time,
evincing in characteristic cusps in $g(t)$ or one of its time-derivatives.
These cusps are followed by zeros of ${\cal L}(t)$, known in statistical physics as Fisher zeros of the
partition function~\cite{Fisher1967}.
In this section we search both pure and mixed state Floquet DQPTs in proposed time-dependent
Hamiltonian Eq.~(\ref{eq1}) to study features of DQPTs in the quantum Floquet systems.

\subsection{Pure state dynamical topological quantum phase transition}
A straightforward calculation yields the exact expression of the Loschmidt amplitude  corresponding to the ground state of the proposed model as follows
%
\begin{equation}
\bl
\label{eq11}
{\cal L}(t)=\langle\psi^{-}(0)|\psi^{-}(t)\rangle=\Pi_{k}{\cal L}_{k}(t),
\el
\end{equation}
with
\begin{equation}
\bl
{\cal L}_{k}(t)
=&\langle\chi^{-}_{k}|\psi^{-}_{k}(t)\rangle=e^{-{\it i}\varepsilon^{-}_{k}t}\langle\chi^{-}_{k}|U_{R}(t)|\chi^{-}_{k}\rangle
\\
\no
=&
\left[\frac{h_{xy}^{2}(k)
+
\eta^{2}_z(k) e^{{\it i}\omega t}
}{
h_{xy}^{2}(k)
+\eta^{2}_z(k)
}\right]
e^{-i\varepsilon^{-}_{k}t}
.
\el
\end{equation}
%
Analysing Eq.~(\ref{eq11}) reveal that the zeros of ${\cal L}(t)$ at which DQPTs occur, take place at critical times  
%
\begin{equation}
t^{\ast}_{n}=(2n+1)\frac{\pi}{\omega}=(n+\frac{1}{2}) t^{\ast}
;\;\;\;
t^{\ast}=2\pi/\omega,~~n{\cal{2}}\mathbb{Z}
\label{eq12},
\end{equation}
%
only whenever there is a mode $k^{\ast}$ that satisfies $h_{z}(k^{\ast})=\omega$.
Also, a critical mode $k_{c}$ exists when $\cos(k^{\ast})=(\omega+\mu)/{\cal W}$, and results $\omega_{1}\leqslant\omega\leqslant\omega_{2}$,
with $\omega_{1}=-{\cal W}-\mu$, and $\omega_{2}={\cal W}-\mu$.
%
\begin{figure*}[]
\centerline{\includegraphics[width=0.71\textwidth]{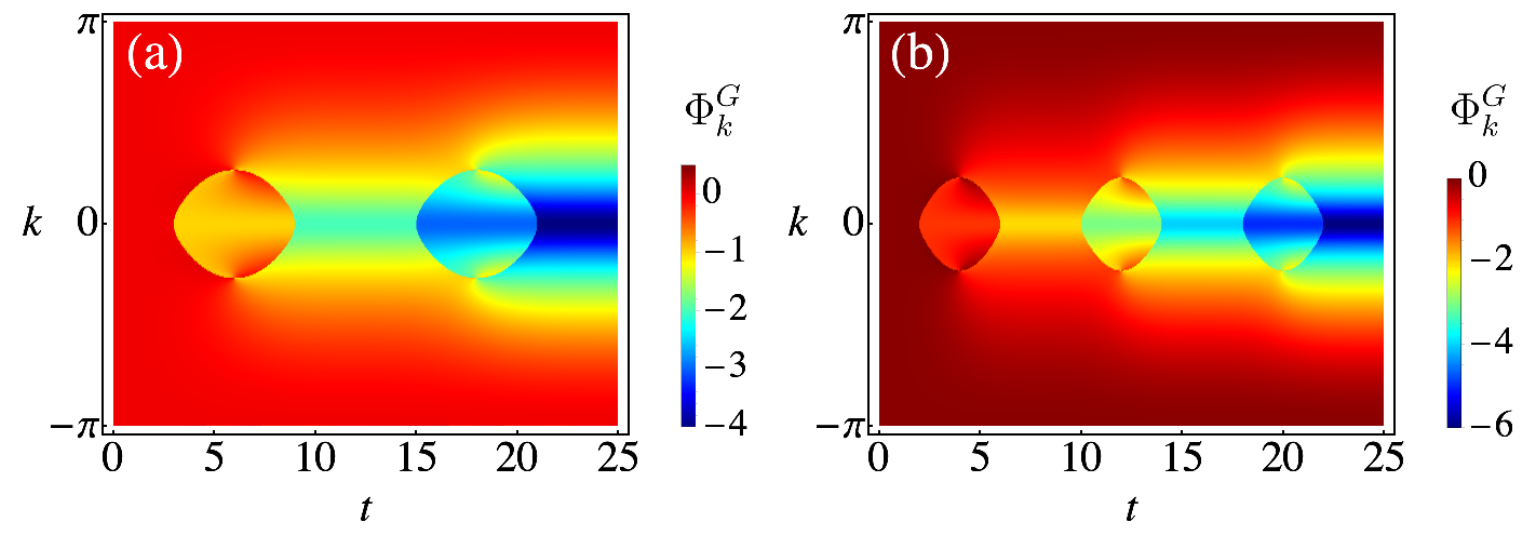}
\hspace{0.2cm}
\includegraphics[width=0.61\columnwidth]{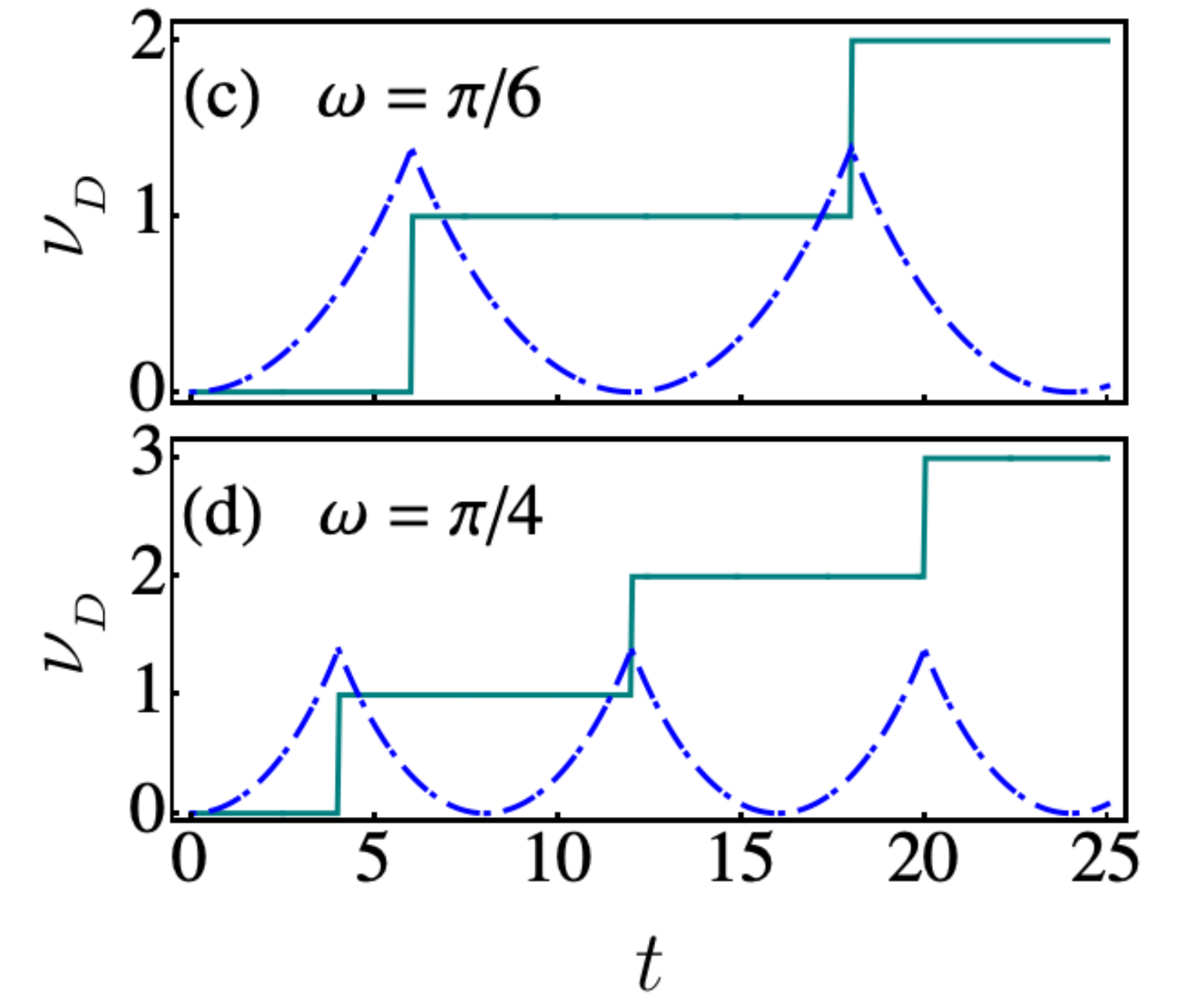}}
\vspace{-0.35cm}
\caption{(Color online) The density plot of geometric phase versus $k$ and $t$ for (a) $\omega=\pi/6$ and (b) $\pi/4$.
The dynamical topological order parameter versus time for: (c)  $\omega=\pi/6$, and  (d) $\omega=\pi/4$ [dashed lines show the dynamical free energy $g(t)$ versus $t$]. The Hamiltonian parameters are set as ${\cal W}=\Delta=\pi$ and $\mu=\pi/2$.}
\label{fig2}
\end{figure*}
The condition $h_{z}(k)=\omega$ is equivalent to the resonance characteristic in the Schwinger-Rabi model of a quasi-spin in a rotating effective magnetic field~\cite{Schwinger1937}.
Thus, two regimes emerge in the proposed time-dependent Hamiltonian Eq.~(\ref{eq1}). First, the resonance regime where the probability of the quasi-spins flip becomes the maximum possible value. Second, the non-resonance regime where the quasi-spin population does not completely cycle the population between the spin down and up states.
We should stress that DQPTs region coincides with the adiabatic regime where the quasi-spins trace the time-dependent effective magnetic field.
While in the region where quasi-spins in a rotating magnetic field feel a constant effective Zeeman field no DQPTs occur~\cite{Zamani2020}.

The density plot of Loschmidt echo (squared modulus of LA), $R_{k}(t)$, and dynamical free energy, $g(t)$,  are shown in Figs.~\ref{fig1}(a-c).
It is clear that, in the resonance regime [Figs.~\ref{fig1}(a-b)] there exist critical points $k^{\ast}$ and $t^{\ast}$, where ${\cal L}_{k^{\ast}}(t^{\ast})$ becomes zero.
In contrast, there is no such critical point in a non-resonance regime.
As shown in Fig.~\ref{fig1}(c), the non-analyticities in the dynamical free energy and DQPT,  arise for the driving frequency at which the quasi-spins are in the resonance situation.

As stated in the Introduction, a dynamical topological order parameter has been proposed to indicate the topological features
emerge in DQPTs. The dynamical topological order parameter represents integer values as a function of time and shows unit magnitude jumps at the critical times at which the DQPTs appear. The dynamical topological order parameter is given~\cite{budich2016dynamical}
%
\begin{eqnarray}
\label{eq13}
\nu_D(t)=\frac{1}{2\pi}\int_0^\pi\frac{\partial\phi^G(k,t)}{\partial k}\mathrm{d}k,
\end{eqnarray}
%
where the geometric phase $\phi^G(k,t)$ is gained from the total phase $\phi(k,t)$ by subtracting the dynamical
phase
$\phi^{D}(k,t)$:
 $\phi^G(k,t)=\phi(k,t)-\phi^{D}(k,t)$.
The total phase $\phi(k,t)$ is the phase factor of LA in its polar coordinates representation,
i.e.,
$${\cal L}_{k}(t)=|{\cal L}_{k}(t)|e^{i\phi(k,t)},$$
and
$$\phi^{D}(k,t)=-\int_0^t \langle \psi_{k}^{-}(t')|H^{F}_{k}|\psi_{k}^{-}(t')\rangle dt',$$
in which $\phi(k,t)$ and  $\phi^{D}(k,t)$ can be calculated  as follows
%
\begin{eqnarray}
\label{eq14}
\bl
&\phi(k,t)=
-\varepsilon^{-}_{k}t
+
 \tan^{-1}\Big(\frac{
\eta^{2}_z(k)
 \sin(\omega t)}{h_{xy}^{2}(k)
 +
\eta^{2}_z(k)
\cos(\omega t)}\Big),
\hspace{.71cm}
\\
&
\phi^{D}(k,t)=
\Big[
\frac{
h_{xy}^{2}(k)+h_{z}(k)
[h_{z}(k)-\omega]
}{
2\Omega_{R}
}
\Big]t.
\el
\end{eqnarray}
%
%
The geometric phase $\phi^G(k,t)$ and $\nu_{D}(t)$ have been illustrated in Figs.~\ref{fig2}(a-c)
for different values of the driving frequencies in the resonance regime, showing excellent agreement with the
analytical result. The plots display singular changes in successive critical times $t_{n}^{\ast}$ at critical momentum $k^{\ast}$,
where characterizes the topological aspects of DQPTs.

%
\begin{figure*}[t]
\centerline{\includegraphics[width=\textwidth]{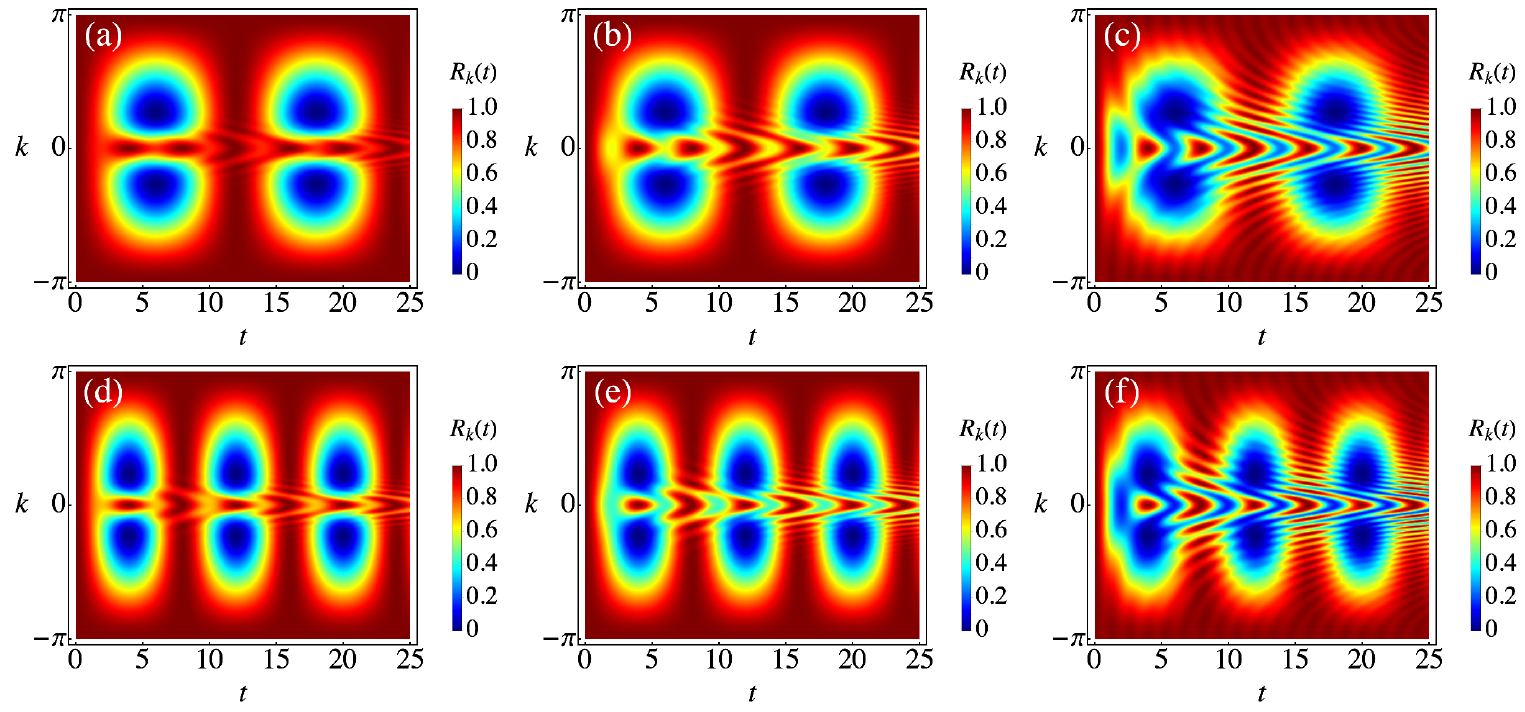}}
\vspace{-0.35cm}
\caption{(Color online) The Loschmidt echo of mixed state versus $k$ and $t$ for
(a) $\omega=\pi/6$, and $\beta=1$,
(b) $\omega=\pi/6$, and  $\beta=2$,
(c) $\omega=\pi/6$, and $\beta=3$,
(d) $\omega=\pi/4$, and $\beta=1$,
(e) $\omega=\pi/4$, and $\beta=2$,
and
(e) $\omega=\pi/4$, and $\beta=3$. Here we set ${\cal W}=\Delta=\pi$ and $\mu=\pi/2$.}
\label{fig3}
\end{figure*}
%

%
\begin{figure}[t]
\centerline{\includegraphics[width=0.95\linewidth]{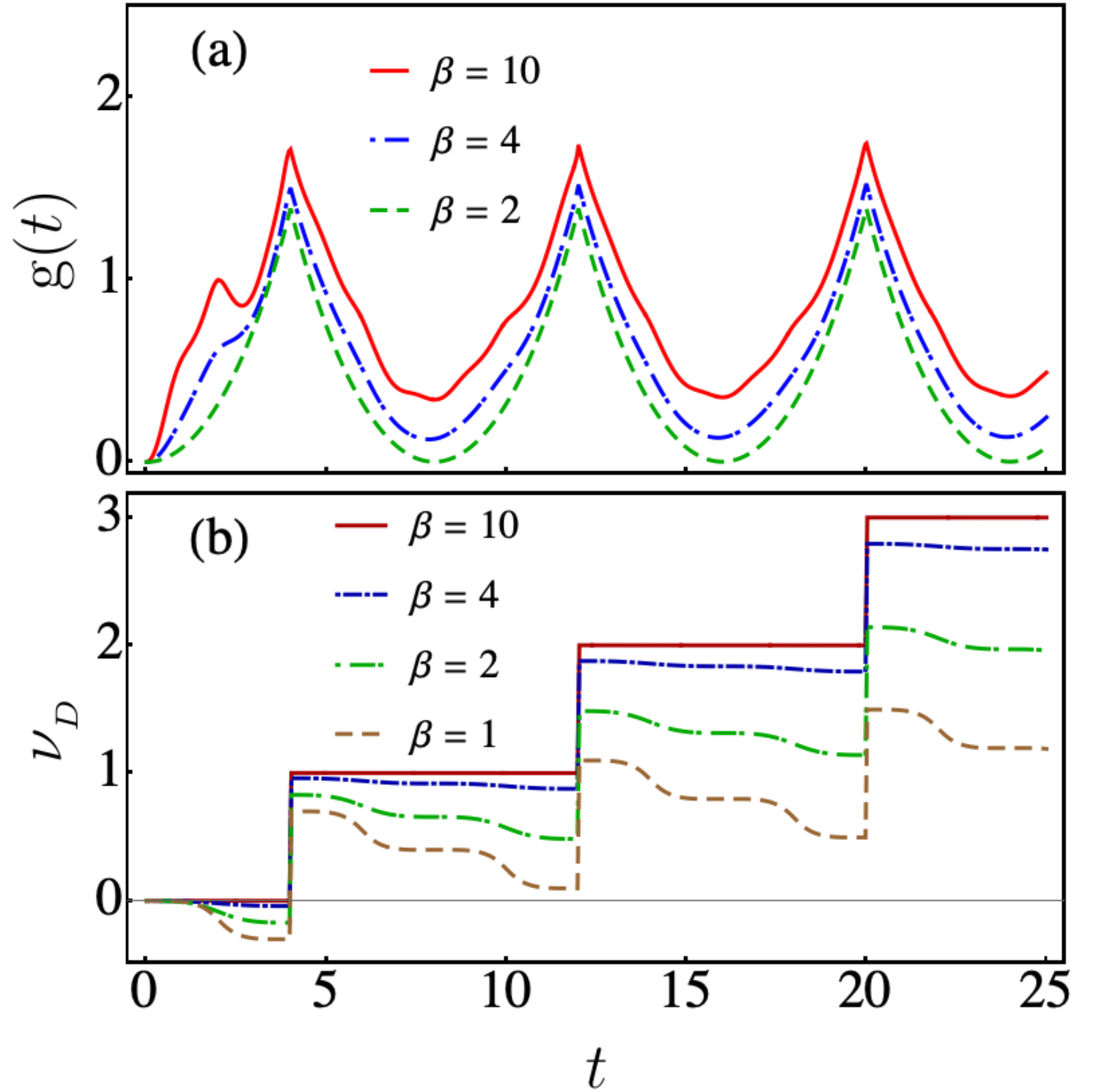}
}
\vspace{-0.15cm}
\caption{(Color online) (a) The dynamical free energy of mixed state $g(t)$
versus time for different values of $\beta$. (b) The mixed state dynamical
topological order parameter versus time for different values of $\beta$ and
$\omega=\pi/4$. The Hamiltonian parameters are set ${\cal W}=\Delta=\pi$ and $\mu=\pi/2$.}
\label{fig4}
\end{figure}
%

\subsection{Mixed state dynamical topological phase transition}
In experiments~\cite{flaschner2018observation,jurcevic2017direct}, the initial state of far-from-equilibrium, which system is prepared, is usually not a pure state but rather a mixed state.
Consequently, on the theoretical front, generalized Loschmidt amplitude (GLA) for mixed thermal states has been established, which perfectly yields the non-analyticities appeared in the pure state DQPTs~\cite{Bhattacharya,Heyl}.
Now, we study the mixed state Floquet DQPTs in the time-dependent Hamiltonian, Eq.~(\ref{eq1}). The GLA for thermal mixed state is described by
%
\begin{equation}
{\cal GL}(t) =\prod_{k} {\cal GL}_{k}(t)=
\prod_{k}
{\rm Tr}
 \Big[
 \rho_{k}(0) U(t)
 \Big],
\label{eq15}
\end{equation}
%
%
where $\rho_{k}(0)$ is the mixed state density matrix at time $t=0$, and $U(t)$ is the time-evolution operator.
The time-evolution operator and the mixed state density matrix of Hamiltonian in Eq.~(\ref{eq1}) are given by
%
\begin{equation}
\label{eq16}
U(t)=U_{R}(t)e^{-{\it i}H^{F}_{k}t}=e^{{\it i}\omega( \sigma_{0} -\sigma^{z})t/2}e^{-{\it i}H^{F}_{k}t},
\end{equation}
%
and
%
\begin{equation}
\label{eq17}
\rho_{k}(0)=\frac{e^{-\beta H^{F}_{k}}}{{\rm Tr}(e^{-\beta H^{F}_{k}})}
=\frac{1}{2}
\Big[
 \sigma_{0} -\tanh(\frac{\beta\Delta_{k}}{2}){{\hat n}_k}\cdot {\vec {\sigma}}
\Big],
\end{equation}
%
respectively.
Here, $H^{k}_{k}=\frac{1}{2}(\omega \sigma_{0} +\Delta_{k}{\hat n}_k\cdot\vec {\sigma})$
with $\Delta_{k}=|\varepsilon^{+}_{k}-\varepsilon^{-}_{k}|$,
${\hat n}_k=[0,-h_{xy}(k),h_{z}(k)-\omega]/\Delta_{k}$ and $\beta=1/T^{}$ is the inverse temperature
with Boltzmann constant $K_{B}=1$.
A rather lengthy calculation results in an exact expression for GLA, which is represented by
%
\begin{equation}
\bl
&
{\cal GL}_{k}(t)=\frac{1}{\Delta_{k}}
\Big[
\Upsilon_1(k,t)+i\Upsilon_2(k,t)
\tanh(\frac{\beta\Delta_{k}}{2})
\Big],
\el
\end{equation}
where $\Upsilon_1(k,t)$ and $\Upsilon_2(k,t)$ are identified as
%
\begin{equation}
\bl
\no
&
\Upsilon_1(k,t)=
\\&
\hspace{0.4cm}
\Delta_{k}\cos(\frac{\omega t}{2})\cos(\frac{\Delta_{k}t}{2})
-
[
h_z(k)-\omega]
\sin(\frac{\omega t}{2})\sin(\frac{\Delta_{k}t}{2});
\\
&
\Upsilon_2(k,t)=
\\
&
\hspace{0.4cm}
\no
\Delta_{k}\cos(\frac{\omega t}{2})\sin(\frac{\Delta_{k}t}{2})
\!+\!
[
h_z(k)
\!-\!
\omega
]
\sin(\frac{\omega t}{2})\cos(\frac{\Delta_{k}t}{2})
.
\el
\end{equation}
The density plot of modulus of GLA has been displayed versus time $t$ and $k$ in Figs.~\ref{fig3}(a-f) for different values of $\beta$ and driving frequencies in resonance regime:
$\omega=\pi/6$ and $\omega=\pi/4$. As seen, the critical points $k^{\ast}$ and $t^{\ast}$, where GLA becomes zero, are exactly the same as the corresponding one in LA.
Therefore, we expect that the mixed state DQPTs appear in the resonance regime even at finite temperatures. The comparison of Fig.~\ref{fig1}(a-b) with
Figs.~\ref{fig3}(a-f) reveals that, GLA deformed versus time. Our numerical results show that the deformation enhances by increasing the temperature and time.
The dynamical free energy of GLA has been depicted versus time in Fig.~\ref{fig4}(a) for different values of $\beta$ and driving frequency $\omega=\pi/4$.
It can be clearly seen that, GLA correctly captures the critical mode $k^{\ast}$, and critical time $t^{\ast}$, observed during the pure state DQPT, but the height of cusps increases by increasing temperature.
It should be stressed that, as the temperature gets smaller than the effective temperature associated with the minimum energy gap, the critical modes and times of the mixed state DQPT, remain unaffected. For higher temperatures the hallmark of DQPT wiped out, which express a traverse to a the regime without DQPT.
\\

Moreover, for mixed state DQPT topological invariant has been proposed to lay out its topological characteristics.
In the mixed state DQPT the total phase and dynamical phase are given as
$$\phi(k,\beta,t)={\rm Arg}\Big[{\rm Tr}\Big[\rho(k,\beta,0)U(t)\Big]\Big];$$
and
$$\phi^{D}(k,\beta,t)=-\int_{0}^{t} {\rm Tr}\Big[\rho(k,\beta,t')H(k,t')\Big]dt',$$
 respectively.
The  topological invariant $\nu_D(t)$ can be calculated using Eq.~(\ref{eq12}) for mixed state in which
$$\phi^{G}(k,\beta,t)=\phi(k,\beta,t)-\phi^{D}(k,\beta,t).$$
After a lengthy calculation, one can obtain the total phase $\phi(k,\beta,t)$ and the dynamical phase $\phi^{D} (k,\beta,t)$ as follows
%
	\begin{eqnarray}
	\bl
	\no
	\phi(k,\beta,t)=
	&
	{\rm Arg}
	\Big[
	{\rm Tr}
	\Big[
	\rho(k,\beta,0)U(t)
	\Big]
	\Big]
	\\
	=&
	\tan^{-1}
	\Big[
	\Big(
	\frac{
	\Upsilon_2(k,t)
	}
	{
	\Upsilon_1(k,t)
	}
	\Big)
	\tanh(\frac{\beta\Delta_{k}}{2})
	\Big],
	\\
	\label{eq18}
	\phi^{D} (k,\beta,t)
	=&
	-\int_{0}^{t}dt'
	{\rm Tr}
	\Big[
	\rho(k,\beta,t')H(k,t')
	\Big]
	\\
	=&
	\tanh(
	\frac{ \beta\Delta_{k} }{2})
	\Big[
	\frac{ h_{z}(k)
	[h_{z}(k)-\omega]
	+h_{xy}^2(k)
	}{2\Delta_{k}}
	\Big]t.
	\el
	\\
	\end{eqnarray}
%
%
In Fig.~\ref{fig4}(b) the mixed state topological invariant has been plotted for driving frequencies $\omega=\pi/4$ and different values of $\beta$.
It can be seen clearly that $\nu_D(t)$ exhibits a nearly perfect quantization (unit jump) as a function of time between the two DQPTs times.
When temperature is smaller than the effective temperature, associated with the minimum energy gap, the quantized structure of $\nu_D(t)$
can be observed.
Although sudden jumps of $\nu_D(t)$ is present at higher temperatures, it does not show a quantized value to
display a topological character. Consequently, mixed state DQPT exist and are signaled by nearly quantized mixed state dynamical topological order parameter, when the temperature is below a crossover temperature.

\section{Entanglement}
As stated, characterization of quantum phase transitions (QPTs) and quantum phases via purity entanglement
measures~\cite{Barnum2003,Barnum2004,Somma2004,Batle2015} and ES~\cite{Gong2018,Li2008,Chang2020,Stojanovi2020,Chang2020,Lu2019}
is one of the most intriguing research topics in condensed-matter physics~\cite{Amico2008}.
In this section we study the purity entanglement measure and entanglement spectrum as a generalization of entanglement
in the time dependent Hamiltonian Eq.~(\ref{eq1}).
We show that both purity entanglement measure and entanglement spectrum can detect the boundary of the driven frequency range over which
DQPTs take place.

\subsection{Entanglement spectrum}
In the following, we focus on the entanglement spectrum of the proposed time-dependent Hamiltonian Eq.~(\ref{eq1}).
To calculate it, 
we should obtain two $l\times l$ correlation matrices
$C$ and $F$ with the matrix elements
$C_{mn}=\langle\psi(t)\rvert c_m^{\dagger} c_n\lvert \psi(t)\rangle$
and $F_{mn}=\langle\psi(t)\rvert c_m^{\dagger} c_n^{\dagger}\lvert \psi(t)\rangle$,
respectively.
Here $ 1 \leq m,n\leq l $, and
 entanglement spectrum can be obtained from
$2l \times 2l$ correlation matrix defined as,
%
\begin{equation}
	\bold{\cal{C}}_l (t) = \left({\begin{array}{cc} I-C & F \\ F^{\dagger} & C \end{array}}\right),
	\label{eq_19}
\end{equation}
%
where $I$ is the $l\times l$ identity matrix.
The single-particle entanglement spectrum obtained  by 
 the eigenvalues of the correlation matrix $\bold{\cal{C}}_l (t)$~\cite{Hughes2011}, and
they come in pairs of $\xi_{m}(t)$ and $1-\xi_{m}(t)$~\cite{Su2020}. 
Moreover, the entanglement entropy of the sub-block of size $l$ is given by,
%
\begin{equation}
\no
S_l(t)=-{\rm Tr}\left[\bold{\cal{C}}_l (t) \log \bold{\cal{C}}_l (t)\right]=-\sum_{m}^{2l} \xi_{m}(t) \log \left[\xi_{m}(t)\right].
\end{equation}
%
Having obtained the time evolved state in Eq.~(\ref{eq9}) the correlation matrix elements can be calculated as follows
%
\begin{eqnarray}
\no
	C_{mn} &=& \frac{1}{L} \sum_k \lvert v_k(t,\omega) \rvert^2 e^{-ik(m-n)}, \\
\no
	F_{mn} &=& \frac{1}{L} \sum_k  v_k^*(t,\omega) u_k(t,\omega) e^{-ik(m-n)}.
\end{eqnarray}
%
The knowledge of the correlation matrix $\mathcal{C}_l(t)$ enables us to calculate the entanglement spectrum.
We have calculated the eigenvalues of the correlation matrix for $l=40$. Our numerical calculation shows that the eigenvalues of the correlation matrix are time-independent. In addition the derivative of all the eigenvalues with respect to the driven frequency show divergence at the boundary of the resonance regime where DQPTs happen.

\begin{figure}[]
\centerline{\includegraphics[width=1\linewidth]{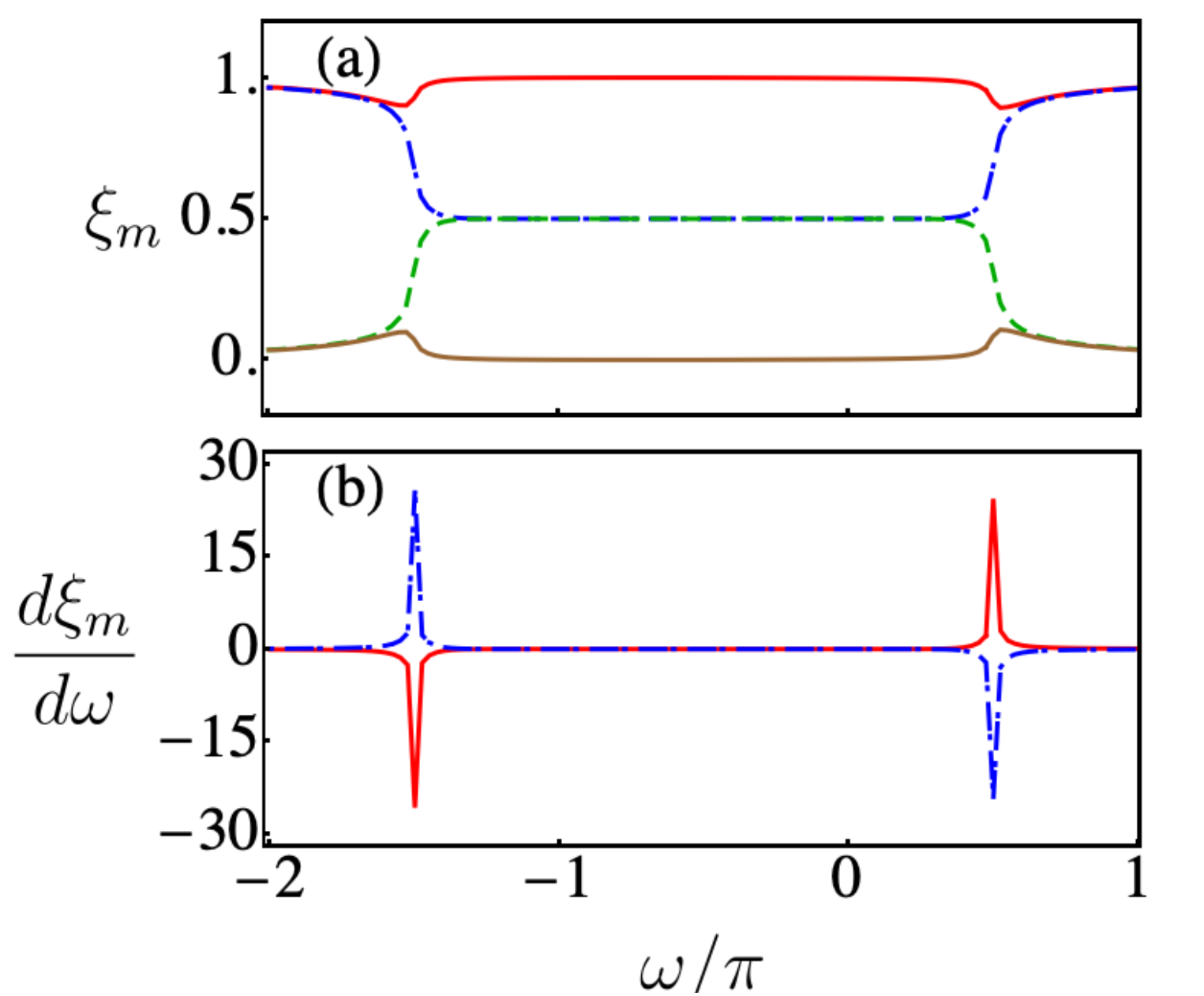}
}
\vspace{-0.15cm}
\caption{(Color online)
(a) The four middle eigenvalues of
entanglement spectrum versus $\omega$.
(b) The derivative of two middle eigenvalues of entanglement spectrum
with respect to driven frequency versus $\omega$. We set ${\cal W}=\Delta=\pi$ and $\mu=\pi/2$.
}
\label{fig5}
\end{figure}
%

The four middle eigenvalues of correlation matrix are shown in
Fig.~\ref{fig5}(a), and
the derivative of two middle eigenvalues of correlation matrix have been plotted in Fig.~\ref{fig5}(b) for
${\cal W}=\pi$ and $\mu=\pi/2$. As seen, the boundary resonance region i.e., $\omega_{1}=-3\pi/2$, and $\omega_{2}=\pi/2$ have been signaled
by the derivative of the eigenvalues with respect to the driven frequency.
 As evidence, two middle eigenvalues of the correlation matrix are degenerate at the resonance region.
This phenomenon is similar to what happened in the entanglement spectrum crossing~\cite{Gong2018,Canovi2014}.
In the entanglement spectrum crossing phenomena, the topological phase results in degeneracies of low-lying entanglement spectrum~\cite{Fidkowski2010}.
In other words, the low-lying entanglement spectrum will be $1/2$ in the topological phase.
Thus, the entanglement spectrum is able to detect the topological phase i.e., the resonance region where dynamically is topological.
Further, it is noteworthy to mention that, the entanglement entropy is zero in the non-resonance regime and is one in the resonance regime. This means the system at the resonance regime, where dynamically is topological, is entangled while it is disentangled in non-resonance region.

\subsection{Purity entanglement measure}
\begin{figure}[]
\centerline{\includegraphics[width=1.04\linewidth]{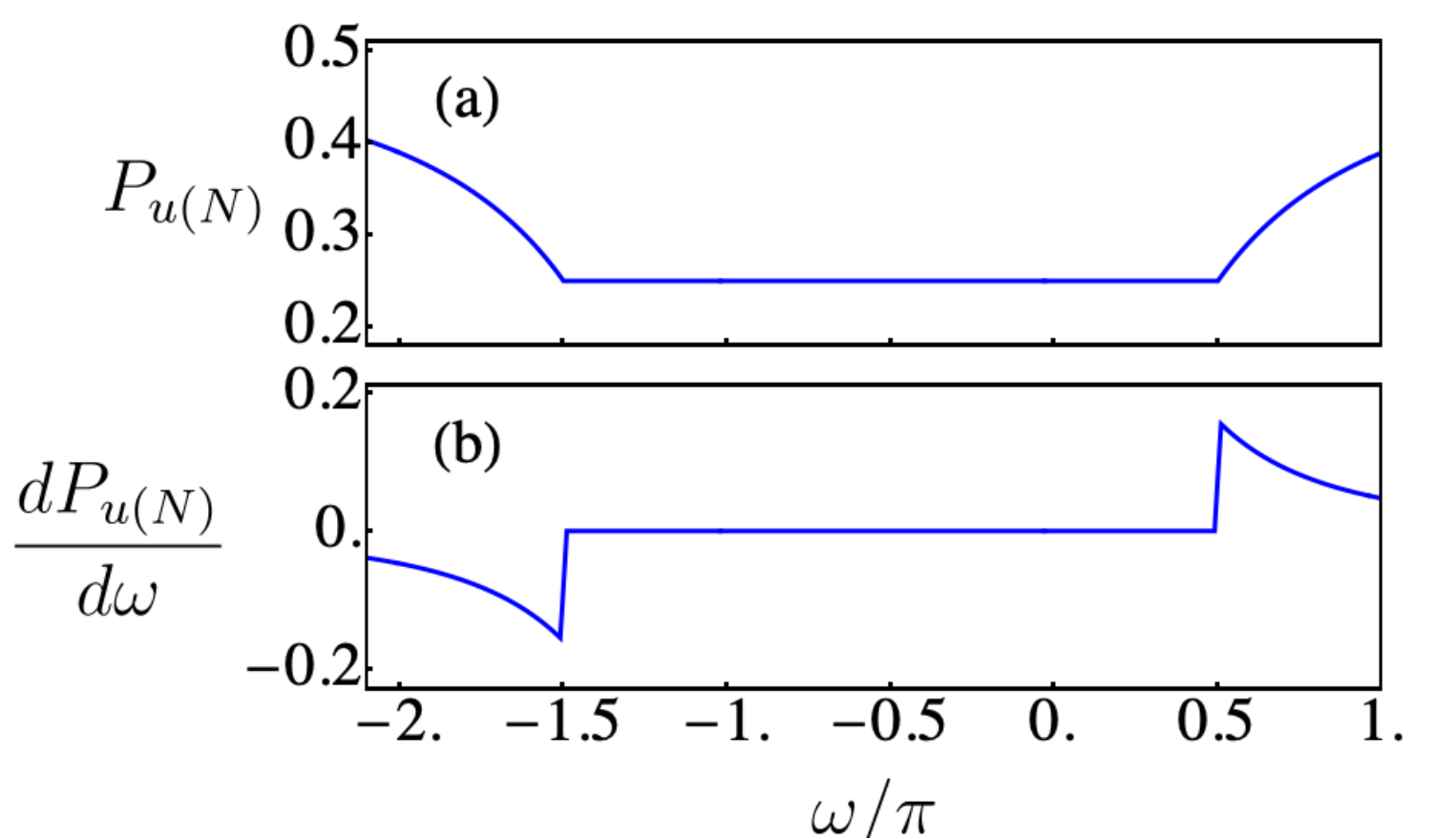}
}
\vspace{-0.35cm}
\caption{(Color online)
(a) The purity entanglement measure and (b) its derivative
with respect to driven frequency  versus $\omega$,  for ${\cal W}=\Delta=\pi$ and $\mu=\pi/2$.}
\label{fig6}
\end{figure}

As it is introduced in Ref.~[\onlinecite{Somma2004}],
the purity entanglement measure, $u(N)$ purity, is a good measure of generalized entanglement to capture the phase transition in the XY model in a transverse field.
When the ground state of the system is unentangled the purity is one, while zero purity means the ground state of the system is fully-entangled.
Moreover, the properties of the $u(N)$ purity has been investigated in Ref.~[\onlinecite{Batle2015}] for the XY in the presence of a time-dependent magnetic field, and show that this measure still
captures the relevant correlations of the system and gives information about the physics underlying the evolution. Now, following the route provided in Ref.~[\onlinecite{Batle2015}], the $u(N)$ purity of the time evolved state in Eq.~(\ref{eq9}) is given as
%
\begin{equation}
\label{eq20}
P_{u(N)}=\frac{2}{\pi}\int_{0}^{\pi}\Big(v_{k}(t,\omega)v_{k}^{\ast}(t,\omega)-\frac{1}{2}\Big)^{2}dk.
\end{equation}
%
Our calculation shows that the purity measure of a state, Eq.~(\ref{eq9}), is time-independent which is plotted in Fig.~\ref{fig6}(a) versus driven frequency for Hamiltonian parameters ${\cal W}=\pi$ and $\mu=\pi/2$.
As reflected, the purity measure, in the resonance regime, has a non-zero constant value but is less than one which means the time evolved ground state of the system is entangled. In the non-resonance region, the purity measure goes to one as the absolute value of driven frequency increases.
In such a case, the ground state of the time-dependent Hamiltonian Eq.~(\ref{eq1}) is unentangled.
The corresponding derivative of purity measure with respect to the driven frequency is also plotted in Fig.~\ref{fig6}(b) versus driven frequency. As is clear, the derivative of the purity measure shows a discontinuity at the boundary of the resonance region.
Therefore, the derivative of the purity measure can truly capture the boundary of the resonance region where DQPTs occur.

%
%

\section{Conclusion}
We have investigated both pure and mixed states Floquet dynamical
quantum phase transition in the one dimensional $p$-wave superconductor with a time-driven pairing phase.
The proposed time-dependent fermions system is equivalent to noninteracting quasi-spins imposed by a time-dependent effective magnetic field in Fourier space.
We have shown that there exists a range of driven frequency over which the quasi-spins are resonance.
In the resonance region, the population of spin down and up states completely is a cycle and both states can be completely populated. While in the non-resonance regime spin-flip (Rabi transition) probability is less than the maximum possible value, $1$, and the state in which the system is initially prepared never completely depopulated.
We have also shown that there is a range of driving frequency, where dynamical topological quantum phase transitions
appear, without requiring any quantum quenches and that range corresponds to the resonance regime.
Moreover, we study the entanglement spectrum and purity measure entanglement.
The results state that the region, where the Floquet dynamical topological quantum phase transitions arise, signaled by the degeneracy of the entanglement spectrum.
In addition derivative of the entanglement spectrum/purity entanglement measure with respect to the driven frequency shows divergence/discontinuity at the boundary of resonance regime.
%

\section*{ACKNOWLEDGMENTS}
A.A. acknowledges the support of the Max Planck-POSTECH-Hsinchu Center for Complex Phase Materials, and financial support from the National Research Foundation (NRF) funded by the Ministry of Science of Korea (Grant No. 2016K1A4A01922028).
%

\bibliography{REFKC}

\end{document}